\documentstyle[editedvolume]{crckapb} 
 
\begin{opening}
\title{Inner region accretion flows onto black holes}

\author{Menas Kafatos and Prasad Subramanian}
\institute{Center for Earth Observing and Space Research\\
George Mason University\\Fairfax, VA 22030, U.S.A.}
 
\end{opening}

\runningtitle{Accretion flows onto black holes}
 
\begin{document}
 
 
\section{Abstract}
We examine here the inner region accretion flows onto black holes. A variety of models are presented. We also discuss viscosity mechanisms under a variety of circumstances, for standard accretion disks onto galactic black holes and supermassive black holes and hot accretion disks. Relevant work is presented here on unified aspects of disk accretion onto supermassive black holes and the possible coupling of thick disks to beams in the inner regions. We also explore other accretion flow scenarios. We conclude that a variety of scenarios yield high temperatures in the inner flows and that viscosity is likely not higher than alpha $\sim$ 0.01.
 
\section{Introduction}
``Accretion is recognized as a phenomenon of fundamental importance in Astrophysics'' (Frank, King and Raine, 1992). This is indeed the case as gravitational energy released in accretion processes is believed to be the dominant source of energy in a variety of high energy galactic compact sources in binary star systems containing white dwarfs, neutron stars and black holes; as well as extragalactic, supermassive black holes (Shapiro and Teukolsky, 1983). Both spherical/quasi-spherical accretion (Bondi, 1952; Frank et al., 1992 and references therein; Treves, Maraschi and Abramowicz, 1989 and papers therein); and disk accretion (Pringle, 1981; Dermott, Hunter, and Wilson, 1992; Frank et al., 1992 and references therein; Treves et al., 1989 and papers therein) operate and the type of accretion may depend on boundary conditions such as the motion of gas at infinity, its angular momentum per unit mass, etc.
The field of accretion astrophysics is obviously vast. In this paper we concentrate on the inner regions of accretion flows onto galactic (stellar) black hole candidates (GBH) and supermassive black holes (SMBH) in active galactic nuclei (AGN).
Accretion can be a very efficient process. The luminous energy released as matter accretes with mass accretion rate $\dot{M}$ is
{\begin{eqnarray}
\nonumber
L \sim 10^{34} \, (\dot{M}/10^{-9} M_{\odot} \, {\rm yr}^{-1}) \,  {\rm
erg/sec} \, \, \, \, {\rm for \, a \, white \, dwarf} \\  
L \sim 10^{37} \, (\dot{M}/10^{-9} M_{\odot} \, {\rm yr}^{-1}) \,  {\rm
erg/sec} \, \, \, \, {\rm for \, a \, neutron \, star \, or \, a \, black \, hole}
\end{eqnarray}}
 
With corresponding efficiencies ($L = \eta \dot{M}\, c^2$)

{\begin{eqnarray}
\nonumber
\eta \sim 10^{-4} \, \, \, \, \, {\rm for \, a \, white \, dwarf} \\
\nonumber
\eta \sim 0.05 \, \, \, \, \, {\rm for \, a \, neutron \, star}\\
\nonumber
\eta \sim 0.06 \, \, \, \, \, {\rm for \, a \, Schwarzschild \, black \, hole} \\
\eta \sim 0.42 \, \, \, \, \, {\rm for \, a \, maximally \, rotating \, Kerr \, black \, hole}
\end{eqnarray}}

In spherical accretion, the specific angular momentum is, by definition zero. High temperatures can be achieved in the inner regions but whether this radiation escapes or not depends on the optical thickness of the accreting gas (Loeb and Laor 1992). Very high temperatures can be achieved in the inner regions of accretion disks: Minimum values apply in an optically thick gas where LTE is achieved, 

\begin{equation}
\frac{1}{4} \, a \, T_{min}^4 = F(r) = \frac{3}{8} \pi\,  \frac{GM\dot{M}}{r^3} \, ,
\end{equation}
whereas when complete internalization of gravitational energy by an optically thin gas is achieved,

\begin{equation}
kT_{max} \sim \frac{GMm_p}{r_{ms}}									\end{equation}
 
where $r_{ms}$ is the marginally stable radius. $T_{max} \sim$ 150 MeV ($\sim 2 \times 10^{12}$ K) for a Schwarzschild black hole and $\sim$ 750 MeV ($\sim 10^{13} $K) for a Kerr Black hole.
\S 3 provides a brief summary of a variety of accretion models
applicable to BHs. \S 4 covers the physics of accretion as related to the
physics of viscous flows. \S 5 covers the topic of outflows, presumably
originating close to the central object. \S 6 is discussion and
conclusions.
 
\section{Models}
 
Several models pertaining to disk and quasi-spherical accretion flows are presented here.
 
\subsection{STANDARD DISKS}
Disk accretion proceeds via the outward transfer of angular momentum of
the
accreting gas. The seminal work of Shakura and Sunyaev (1973) provided the basic formalism of accretion in what hence came to be known as ``standard'' disks (see also Novikov and Thorne, 1973; Lynden-Bell and Pringle, 1974). Such disks are optically thick, geometrically thin, radiate locally as black bodies (BB) and have a radial dependence $T(r) \sim r^{-3/4}$. T is analogous to the effective temperature of a star. The modified BB spectrum has a characteristic broad peak, at low-frequencies $S_{\nu} \sim \nu^2$ while at high frequencies the spectrum drops exponentially. In between, but over a not too large range of frequencies the spectrum depends on frequency as $\nu^{1/3}$, the characteristic accretion disk law (Pringle, 1981; Kafatos, 1988). For typical parameters applicable to SMBHs, $10^8 M_{\odot}$ and near-Eddington accretion, the peak of the modified black body spectrum occurs in the UV, ${\rm log} \nu!
 \sim 15.3 - 15.5$, with $T \sim 20,000 - 30,000$ K (Ramos, 1997); whereas for GBHs, $\sim 10 M_{\odot}$, the peak occurs below $\sim$ 1 keV, with $T \sim 1 - {\rm a \, few} \times 10^7  $K (Shapiro, Lightman, and Eardley, 1976). 
In reality the disk accretion is more complicated than the above simple expressions. Several regions in the accretion flows have been identified (Shakura and Sunyaev, 1973; Novikov and Thorne, 1973; Kafatos, 1988): i) an outer region where gas pressure dominates over radiation pressure and where the opacity is predominantly free-free; ii) a middle region where gas pressure again dominates but the opacity is primarily due to electron scattering; and iii) an inner region where radiation pressure dominates over gas pressure and the opacity is primarily due to electron scattering. The latter is expected to occur for $r \leq 50 r_g$, where $r_g$ is the gravitational radius of the black hole, $GM/c^2$. The simple thick disk solution applies to the ``outer'' disk region (Novikov and Thorne, 1973) as well as in accretion disks around white dwarfs.

\subsection{MODIFIED DISKS}

Further considerations indicate that optically thick disks have to be modified. Modifications include electron scattering and Comptonization. If electron scattering dominates, the emitted flux is lower (Rybicki and Lightman, 1979) as follows

\begin{equation}
I_{\nu} = B_{\nu} \sqrt{\frac{\kappa_{\rm abs}}{(\kappa_{\rm abs} + \kappa_{\rm es})}}
\end{equation}

Modifications due to electron scattering are important for ${\rm log}
\nu > 15 $ (Ramos, 1997) for SMBHs. Malkan and Sargent (1982), and Ramos
(1997) have applied these modifications. At high energies, photons are
scattered many times by the
electrons before they leave the disk. This was recognized as far back as the original work by Shakura and Sunyaev (1973).  The result is that the (relatively) soft photons emanating from the disk are upscattered by the hot electrons and become hardened. This is known as Comptonization (Shapiro, Lightman, and Eardley, 1976; Sunyaev and Titarchuk, 1980; 1985) and is believed to produce hard, power-law radiation above the usual broad disk peak. Comptonization is important above ${\rm log} \nu = 15.8$ for SMBHs (Ramos, 1997) and above 10 keV for GBH candidates such as Cygnus X-1 (Shapiro, Lightman, and Eardley, 1976).
 
\subsection{TWO-TEMPERATURE DISKS AND ION-SUPPORTED TORI}

If $T_e \sim 10^9$ K in the inner portion of the disk, a two-temperature
solution is obtained (Shapiro, Lightman, and Eardley, 1976: Eilek and
Kafatos, 1983), where the ions are much hotter than the electrons, $T_i
\sim 10^{11} - 10^{13}$K. In such a disk, a puffed-up inner region is
formed supported by the ion pressure. Unsaturated Comptonization transfers energy from the electrons to the soft photons emitted in the cooler, underlying disk. The process is described by the dimensionless parameter y (Shapiro, Lightman, and Eardley, 1976), where y = <fractional energy change per scattering><number of scatterings> or 
 
\begin{equation}
y = (\frac{4kT_e}{m_e c^2}) \, {\rm max} (\tau_{es}, \tau_{es}^2)							\end{equation}

Unsaturated Comptonization occurs for ${\rm y} \sim 1$ and is
appropriate whenever there is a copious source of soft photons in the
inner region (Shapiro, Lightman, and Eardley, 1976; Sunyaev and Titarchuk,
1985). The y-value is related to the energy flux spectral index A (where
the energy flux is measured in ${\rm keV}\, {\rm cm}^-2 \, {\rm s} ^-1\,
{\rm
keV}^-1$) through $ {\rm A} \sim 0.72 \, {\rm y}^{-0.917}$ (Kafatos, 1983) and as such y $\sim$ 1 provides a natural explanation for the spectra of many cosmic sources. The 2T solution is thermally unstable (Piran, 1978) and whether it occurs or not depends on a variety of factors, including whether accretion is at near-Eddington rates, where the Eddington luminosity is ${\rm L}_{\rm Edd} \sim 10^{38} (M/M_{\odot})$ erg/sec. Detailed spectra of 2T disks including relativistic effects, ion-ion collisions and resultant radiation spectra at gamma-rays (from pions and relativistic pairs which subsequently radiate via inverse-Compton)!
 have been calculated by Eilek and Kafatos (1983). Both the Shapiro,
Lightman and Eardley and Eilek and Kafatos solutions apply to
near-Eddington accretion rates. Gamma-gamma scatterings will degrade
gamma-rays above $\sim$ MeV as these photons scatter the softer X-rays emerging from the 2T disk. The resultant optical depth for AGNs (Eilek and Kafatos, 1983) is
 
\begin{equation}
\tau_{\gamma \, \gamma} \sim 5 \times 10^{-2} \, {\rm D}_{\rm Mpc}^2 \,  {\rm E}_{\rm T}\, ({\rm keV}) \, {\rm N} (2{\rm E}_{\rm T}) \, {\rm R}_{\gamma}^{-1} \, ,
\end{equation}

where ${\rm D}_{\rm Mpc}$ is the distance of the AGN in Mpc, ${\rm E}_{\rm T}$ is the relevant threshold X-ray energy for $e^+$ $e^-$ production and N is the corresponding photon flux at the earth (photons ${\rm cm}^{-2}$ ${\rm s}^{-1}$ ${\rm keV}^{-1}$) computed at $2{\rm E}_{\rm T}$. $\gamma$-$\gamma$ scattering will form a broad shoulder $\sim$ 1 MeV with an exponentially-declining tail and the absence of high-energy radiation in many accreting GBH and SMBH (e.g. Seyferts) sources may be explained by this fundamental physical process (Eilek and Kafatos, 1983). High-energy gamma-rays (above 100 MeV - TeV) are probably arising in a jet (see below).
A closely-related model to 2T Comptonized disks is the ion-supported torus model ( Rees, Begelman, Blandord and Phinney, 1982), proposed as the underlying engine in extragalactic jet sources. They found self-consistent 2T solutions
even for $r \gg 2000 r_g$ for sub-Eddington rates as low as $10^{-4} \, \dot{M}_{\rm Edd}$.

\subsection{HOT CORONAE}
Hot coronae may surround an underlying, cooler disk. Corona models have been proposed for Cygnus X-1 (Liang and Price, 1977; Bisnovatyi-Kogan and Blinnikov, 1977) and provide a competing model to hot, 2T disks. Coronae provide a natural explanation for unsaturated Comptonization since a corona would envelop an underlying cool disk where the soft photons emanate. Conduction-balanced coronae (see also Rosner, Tucker, and Vaiana, 1978) would produce temperatures, $T \sim 10^{11}$K, lower than 2T disks but still higher than standard inner accretion disks.
Hot haloes or coronae may also be produced in a bulk motion Comptonization model (see below) and would account for the hard-radiation in such flows.

\subsection{BULK COMPTONIZATION}
A promising theoretical model proposed for the soft-state GBH candidates is the bulk motion Comptonization model (Chakrabarti and Titarchuk, 1995; Titarchuk, Mastichiadis,and Kylafis, 1997; Titarchuk and Zanias, 1998; Shrader and Titarchuk, 1998). In this model (Bautista and Titarchuk, 1999), the production of hard phtons peaks at $\sim 2r_S$ (where $r_S$ is the Schwarzschild radius, = 2 $r_g$). It explains the continuum X-ray spectra of soft-state GBHs in their soft state (Shrader and Titarchuk, 1998). Accretion onto the central BH proceeds via a spherically-converging flow at gravitational free-fall speeds. 
A variant of this model assumes the formation of strong shocks as the convergent inflow speeds become greater than the local sound speed (Chakrabarti and Titarchuk, 1995). In the latter model, the disk is divided into two main components, a standard cool disk which extends to the outer bounday and produces soft (UV) photons; and an optically thin sub-Keplerian halo (or corona) which terminates in a standing shock near the black hole. The post-shock region Comptonizes the soft (UV) photons that are subsequently radiated as a hard spectrum with spectral index $\sim$ 1.5.

\subsection{THICK DISKS AND ADVECTION-DOMINATED FLOWS}
In the above models, it is generally assumed that the disk motion is Keplerian (or quasi-Keplerian). In reality, when radiation pressure is included, the flow becomes non-Keplerian and the disk fattens geometrically (Abramowicz et al.,1978: Jaroszynski et al.,1980; Paczynski and Wiita,1980, Paczynski 1998). A funnel wall is produced as matter cannot reach the axis of rotation. Such thick disk flows may play a role in the production of matter outflows, although the exact mechanism has not been proposed in the above works. These disks are special cases of advective disks (Chakrabarti, 1998).
Besides pressure effects, radial inflow effects have also to be considered. In normal disks, the radial inflow speed is assumed to be negligibly small. In reality, this speed can be large, particularly in the inner regions. As such, the advection term $v dv/dr$ should be included in the momentum equation. It may also be the case that transonic flows always result in accretion onto black holes (Chakrabarti, 1990; Kafatos and Yang, 1994) as the inflow speed has to smoothly join the subsonic regime with the supersonic regime near the horizon. It may also be the case that shocks are prevalent (Chakrabarti, 1990; Yang and Kafatos, 1995; Chakrabarti and Titarchuk, 1995; Chakrabarti, 1996).
Advection-dominated accretion flows (ADAFs) have been widely discussed
in the literature (Narayan and Yi, 1994; Liang and Narayan, 1997; Esin,
McClintock, and Narayan, 1997; Narayan, Mahadevan and Quataert, 1999). At large or super-Eddington rates (Katz, 1977; Abramowicz et al., 1988) optically thick advection solutions are obtained. In these solutions, the large optical depth of the inflowing gas traps or advects it into the central black hole. At low, sub-Eddington accretion rates (Rees et al., 1982; Narayan and Yi, 1994), optically-thin advection flows, termed ADAFs, result. In this model, the accreting gas has low density, is unable to radiate and viscous energy is advected onto the central BH. Optically-thin ADAFs are hot, 2T flows. This model assumes a self-similar solution and can (according to Narayan and co-workers) only operate at low accretion rates and high viscosity values, $\alpha \sim 0.1$. Whether these conditions can be satisfied in realistic flows is another matter.
 
\subsection{ADVECTION DOMINATED INFLOW-OUTFLOW SOLUTIONS}
ADAFs have the generic drawback of having a positive Bernoulli parameter in
the disk. Such accretion flows thus tend to evaporate, before they accrete. Recently, advection dominated inflow outflow solutions (ADIOS) have been proposed (Blandford \& Begelman, 1999) which overcome this drawback by postulating a powerful outflow that carries away excess mass, energy and momentum, thus allowing the accretion to proceed. Subramanian et al. (2000) are investigating a scenario where relativistic outflows can be produced in such an advection-dominated accretion flow, as a result of Fermi acceleration due to collisions with kinks in the tangled magnetic field embedded in the accretion flow. This mechanism has been explored in detail by Subramanian et al. (1999), and it is expected that the low-density environment of advection-dominated flows will be ideal sites for the launching of outflows via this mechanism. On the other hand, time-dependent treatments of quasi-spherical accretion (Ogilvie, 1999) suggest that advection-dominated accretion can proceed without the need for outflows.

\section{Viscosity Mechanisms}

Viscosity in accretion disks has been the subject of investigation for nearly 20 years (for a review, see Pringle 1981). It was recognized early on that ordinary molecular visocisty cannot produce the level of angular momentum transport required to provide accretion rates commensurate with observed levels of emission. In lieu of a detailed model of microphysical viscosity, Shakura \& Sunyaev (1973) embodied all the unknown microphysics of viscosity into a single parameter $\alpha$ according to the prescription
\begin{equation}
t_{r\, \phi} = \alpha \, P \, ,
\end{equation}
where $t_{r\, \phi}$ denotes the $r\, \phi$ component of the viscous stress and
$P$ is the ambient pressure in the disk. Much of the subsequent developments
concentrated on obtaining estimates of the $\alpha$ parameter due to fluid
turbulence (Shakura \& Sunyaev 1973; Goldman \& Wandel 1995) magnetic viscosity
(Eardley \& Lightman 1975; Balbus \& Hawley 1998 and references therein) and 
radiation viscosity (Loeb \& Laor 1992) and ion viscosity (Paczynski 1978; Kafatos 1988). Subramanian et al. (1996) have shown that ``hybrid'' viscosity (neither pure ion visocisty nor pure magnetic viscosity) due to hot ions scattering off kinks in the tangled magnetic field embedded in accretion disks is the dominant form of viscosity in hot, two-temperature accretion disks. 
This work assumes the magnetic field embedded in the accretion disk to
be isotropically tangled. Recent simulations (see, for instance, Armitage
1998) suggest, however, that the manner in which the magnetic field is
tangled might be significantly anisotropic. Based on the detailed
calculations presented in Subramanian et al. (1996), we believe that this
will merely introduce a direction dependece in the hybrid viscosity, but
will not change the overall conclusion that this (hybrid viscosity) is the
dominant form of viscosity in hot accretion disks. The viscosity obtained
is characterized by a viscosity parameter $\alpha \sim 0.01$ (Subramanian 
et al. (1996). In the turbulent viscosity mechanism characterized by
convection (Yang, 1999) when gas pressure dominates, an upper limit to
$\alpha$ of 0.01 is also obtained. For radiative viscosity (Loeb and Laor,
1992) in which radiation pressure dominates, $\alpha$ is greater than 1.
These models are, however, producing optically thick, relatively cool
inner regions.

\section{Jets/Outflows}
Several objects for which much of the preceding discussion of accretion disks
is relevant also exhibit jets/outflows. Although several mechanisms for
producing outflows have been proposed, none of them, with the exception of a few (Das 1998) make the connection between the outflows and the underlying accretion disk. Subramanian et al. (1999) have proposed a model in which the outflow is powered by Fermi acceleration of seed protons due to collisions with magnetic scattering centers embedded in the accretion disk/corona. This is natural mechanism that is expected to operate in any accretion disk, and Subramanian et al. (2000) are examining its viability in the context of the ADIOS scenario. The physical scenario in which the Fermi acceleration of protons (which powers the outflow at its base) takes place is the same as that in which hybrid viscosity (Subramanian et al. 1996) is operative. The energy inherent in the shear flow (gravitational potential energy) is dissipated partly by viscous heating of the thermal protons, and partly by Fermi acceleration of the supra-thermal protons, which in turn form a relativistic outflow.

\section{Discussion and Conclusions}

We have seen that a variety of scenarios predict hot, inner regions,
either 2T, hot coronae or ADAFs, where $T_{max} \sim 10^{11} - 10^{13}
$K. In most models, these hot inner regions occur for radii not much
greater than 10's of gravitational radii. In ADAFs, however, high
temperatures persist for hundreds or even 1,000's of radii. Besides
spectral observations that would reveal hard X-rays and $\gamma$ rays,
timing observations would be crucial (with characteristic timescales
comparable 
to the light-travel time through the hot region). Several of these models
face theoretical difficulties or inconsistencies: 2T disks are unstable
(although appropriate viscosity laws may mitigate this difficulty); hot
coronae are attractive but the mechanism of heating the corona is unknown.
In many (all?) cases, transonic flows and even shocks may be prevalent,
breaking up the ususal assumptions of quasi-Keplerian or even steady disk
structure. ADAFs are particularly problematic: The assumption of
self-similarity is probably erroneous. ADAFs assume that the velocity in
the $\theta$ direction is zero. However, at the boundary between the thin
disk and the geometrically thick ADAF solution, $v_{\theta}$ is not zero.
Also, at the axis this velocity would be zero but then a funnel would be
formed. ADAFs assume no shocks but shocks, transonic flows, etc. are
probably prevalent. Although the usual ADAF assumption of bremmstrahlung
cooling produces ineffient cooling, there is no reason that other more
efficient coolings processes (such as Compton processes) would not be 
operating. Finally, ADAFs seem to require $\alpha \sim 0.1$ whereas
realistic physical calculations for hot, optically thin flows suggest
$\alpha \sim 0.01$.

\end{document}